\documentclass[fleqn,10pt]{wlscirep}
\usepackage[utf8]{inputenc}
\usepackage[T1]{fontenc}

\usepackage{tabularx}
\usepackage{indentfirst}

\usepackage{graphicx}%
\usepackage{multirow}%
\usepackage{amsmath,amssymb,amsfonts}%
\usepackage{amsthm}%
\usepackage{mathrsfs}%
\usepackage{xcolor}%

\newcommand{\ket}[1]{|{#1}\rangle}
\newcommand{\bra}[1]{\langle{#1}|}

\newcommand{\CNOT}{\text{CNOT}}

\title{New Design of three-qubit system \\
with three transmons and a single fixed-frequency resonator coupler}

\author[1,*]{Jeongsoo Kang}
\author[1,**]{Chanpyo Kim}
\author[1,***]{Younghun Kim}
\author[1,+]{Younghun Kwon}
\affil[1]{Department of Applied Physics, Hanyang University, Ansan, 15588, South Korea}

\affil[*]{jskang1202@hanyang.ac.kr}
\affil[**]{freezeticket@hanyang.ac.kr}
\affil[***]{hpoqh@hanyang.ac.kr}
\affil[+]{yyhkwon@hanyang.ac.kr}

\keywords{Superconducting Qubits, Transmon, Single Resonator Coupler, Cross-resonance}

\begin{abstract}
  The transmon, which has a short gate time and remarkable scalability, is the most commonly utilized superconducting qubit, based on the Cooper pair box as a qubit or coupler in superconducting quantum computers. Lattice and heavy-hexagon structures are well-known large-scale configurations for transmon-based quantum computers that classical computers cannot simulate. These structures share a common feature: a resonator coupler that connects two transmon qubits. Although significant progress has been made in implementing quantum error correction and quantum computing using quantum error mitigation, fault-tolerant quantum computing remains unachieved due to the inherent vulnerability of these structures. This raises the question of whether the transmon-resonator-transmon structure is the best option for constructing a transmon-based quantum computer. To address this, we demonstrate that the average fidelity of CNOT gates can exceed 0.98 in a structure where a resonator coupler mediates the coupling of three transmon qubits. This result suggests that our novel structure could be a key method for increasing the number of connections among qubits while preserving gate performance in a transmon-based quantum computer.
\end{abstract}
\begin{document}

\flushbottom
\maketitle
% \thispagestyle{empty}

% \noindent Please note: Abbreviations should be introduced at the first mention in the main text – no abbreviations lists. Suggested structure of main text (not enforced) is provided below.

\section*{Introduction}\label{sec:introduction}

Superconducting qubits are among the most promising devices for quantum computing applications. Most superconducting qubits are based on a Josephson junction (JJ)\cite{Josephson62, Josephson64}, which consists of two superconducting regions separated by a thin insulating layer. As a nonlinear inductor, the JJ allows Cooper pairs in the superconducting circuit to tunnel through the JJ insulator barrier. This feature enables the superconducting circuit to behave as a nonlinear resonator, functioning as an artificial atom fabricated in a laboratory. The Cooper-pair box (CPB)\cite{Nakamura99,Namkung22} is one of the fundamental architecture for superconducting qubits, comprising a gate capacitor and a JJ connected to a microwave source. Furthermore, many architectures based on the JJ, such as flux\cite{Chiorescu03}, quantronium\cite{Vion02}, and phase qubits\cite{Martinis02}, have been introduced. Each of these devices can be designed with a unique energy gap, as the potential energy of the JJ includes higher-order terms beyond the quadratic term of the magnetic flux. However, all these superconducting qubits have not shown a multiqubit system performing high-fidelity universal quantum gates.

Transmons\cite{Koch07}, a modern type of CPB, are commonly employed in constructing superconducting quantum computers. They are notable not only for their fast operation and scalability but also for their longer coherence time compared to charge qubits, owing to their robustness against background charge noise\cite{Nakamura02, Astafiev04, Schreier08, Schlor19}.
Transmons are classified by the tunability of their qubit frequency. Fixed-frequency transmons are more robust to the flux noise, causing dephasing error, and utilize microwave pulses to realize two-qubit gates\cite{Chow11,Kirchhoff18,Rigetti10,Sheldon16,Barron20}. Tunable-frequency transmons exploit their tunability to implement single- and two-qubit gates within more shorter time than fixed-frequency transmons\cite{DiCarlo09, Garcia20, Caldwell18, Barends14}.

It is appearant that transmons are superior to the previous superconducting qubits in the point of view of building practical quantum computer. However, the relatively short decoherence time to execute a large quantum circuit and large gate error rate to implement quantum error-correcting code of them is the challenging obstacle to achieve fault-tolerant quantum computing. Recently, novel superconducting qubits, such as a parity-protected qubit\cite{Smith20} and a unimon\cite{Hyyppa22}, have been introduced to overcome these limitations of transmons. However, these qubits have not yet demonstrated sufficient performance in large multiqubit systems, where effective gate operations or measurement methods have not been adequately discussed.
% Nevertheless, they remain competitive structures for fault-tolerant quantum computing.

% We consider exploiting the tunability of the transmon for gate operations and selecting the appropriate coupling structure for qubit-qubit interactions when building a transmon-based quantum computer. There are two types of transmons based on their resonance frequencies' tunability. A fixed-frequency transmon, with the same structure as the CPB, has a single JJ, and its frequency cannot be tuned using external fields. The controlled-NOT gate implemented in the fixed-frequency transmon exhibits characteristics different from those of the conventional CPB method. In charge qubits, the CNOT gate is implemented by applying an offset voltage impulse\cite{Yamamoto03} or a microwave pulse at the codegeneracy point\cite{Paraoanu06}. In contrast, fixed-frequency transmon systems use cross-resonance (CR)\cite{Chow11,Kirchhoff18,Rigetti10,Sheldon16}, enabling the implementation of a controlled phase gate based on a microwave pulse\cite{Barron20}.

% In contrast to a fixed-frequency transmon, a tunable transmon\cite{Koch07} features a loop of two JJs. This loop structure, known as a DC-SQUID\cite{Tesche77}, plays a crucial role in allowing an external magnetic field to manipulate the energy levels of a tunable transmon. The magnetic flux on the DC-SQUID is used to control one or two tunable transmons\cite{DiCarlo09, Garcia20, Caldwell18, Barends14}, significantly reducing the gate time of a flux-based gate compared to a microwave-based gate.

Designing qubit-qubit interactions is also necessary to construct a practical quantum computer. For the transmon-based quantum computer, couplings between qubits are realized by just positioning qubits closely so that the coupling capacitance mediates the connection\cite{Barends13,Barends14,Noguchi20}, or by inserting a coupler such as a resonator, represented by a coplanar waveguide\cite{Majer07,DiCarlo09} or a microwave cavity\cite{Poletto12,Paik16}, or even a transmon\cite{Yan18,Sete21,Shirai23,Marxer23,Goto22}. Implementing time-invariant couplings is a simple method to realize the superconducting multiqubit system. Still, it is not known which coupling method is the best for implementing a superconducting quantum computer.

Although quantum computers based on transmon qubits have shown excellent performance, the most effective structure for high-performance superconducting quantum computers capable of implementing fault-tolerant quantum computing remains unknown. One of the most well-known structures for transmon-based quantum computers is the transmon-resonator-transmon structure, in which a resonator coupler connects two transmon qubits. IBM\cite{Chow21} and Google\cite{Arute19} use this structure for their superconducting quantum computers and have enhanced hardware performance through error mitigation\cite{Kandala19, Temme17} and quantum error correction\cite{Google21} techniques. Despite considerable efforts, both companies have encountered significant hardware challenges. In IBM's hardware, the qubits are arranged in a heavy-hexagon structure, which allows for system scalability with consistent gate fidelity but limits the number of connections between qubits. This limitation results in quantum algorithms executed on IBM hardware requiring many SWAP gates, which reduce the fidelity of the algorithms. Google's hardware is constructed in a lattice structure that supports quantum error correction codes, such as surface\cite{Fowler12} and XZZX\cite{Bonilla21} codes. However, scaling the system is challenging due to uncontrollable interactions between non-nearest-neighbor qubits caused by parasitic crosstalk errors. Google's hardware uses magnetic flux pulses that cannot be applied locally to each qubit on its dense quantum processor, which limits the gate error rate to a certain level. These cases highlight the need for a new structure with greater connectivity among qubits and high-fidelity two-qubit gates implemented using microwave pulses that can be applied locally to the qubits.

Based on this need, we investigate a new structure with increased connectivity between qubits using high-fidelity two-qubit gates. To implement a microwave-controllable system with improved connectivity, we propose a structure in which a single resonator coupler connects three qubits. We design pulses for CNOT gates and present the pulse parameters for CNOT gates implemented in a three-transmon system. Our results show that a single fixed-frequency resonator coupler can mediate more than two transmon qubits while maintaining average two-qubit gate fidelities of 0.98.

\section*{Results}\label{sec:results}
\subsection*{System Configuration}\label{subsec:systemConfig}
\begin{figure}[t!]
  \centering
  \includegraphics[width=\linewidth]{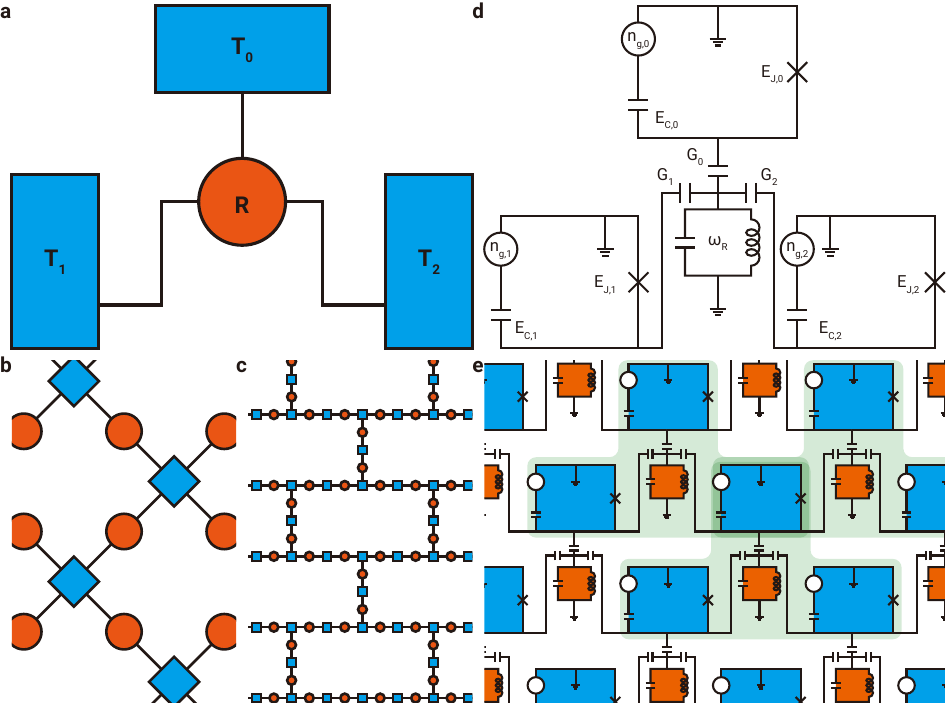}
  \caption{
    Illustrations of the three-transmon system and conventional transmon-based quantum computers. (a) The three-transmon system, which consists of three transmons and a single resonator. T and R denote the transmon qubit (blue box) and the resonator coupler (orange circle), respectively. The black solid line connecting the transmon and the coupler indicates that a coupling capacitor mediates the interaction between the transmon and the coupler. (b) The Google machine (transmon system in a lattice structure\cite{Arute19}). (c) The IBM machine (transmon system in a heavy-hexagon structure\cite{Chow21}). (d) Circuit diagram for the three-transmon system. Each transmon is characterized by its charging energy $E_{C,i}$ and Josephson energy $E_{J,i}$ and is controlled by the gate offset $n_{g,i}$. The resonator has the resonance frequency $\omega_R$. The capacitance of the coupling capacitor, which links the transmon to the resonator, determines the coupling energy $G_i$ of the transmon-resonator interaction. (e) Extended three-transmon system. In this extended structure, each qubit has six nearest-neighbor qubits.
  }
  \label{fig:transmon_systems}
\end{figure}
In this study, we propose a new structure consisting of three transmon qubits and a single resonator coupler, which we refer to as the \textit{three-transmon system}. Figure \ref{fig:transmon_systems}a shows a schematic of the three-transmon system. In this diagram, the blue squares represent the transmon qubits, which are connected by an orange circle, denoting the resonator coupler. The black lines indicate the capacitive qubit-coupler couplings. The coupler in the three-transmon system mediates more qubits than the transmon-resonator-transmon structure. Figures \ref{fig:transmon_systems}b and \ref{fig:transmon_systems}c show the qubit map diagram of the superconducting quantum processor based on the transmon-resonator-transmon structure. We demonstrate that the resonator coupler in our new structure can mediate more qubit-qubit interactions than traditional structures. \ref{fig:transmon_systems}d shows the lumped-element circuit diagram of the three-transmon system. Note that none of the elements in the three-transmon system have a tunable frequency structure using an external flux. This implies that we used only microwave pulses to control the three-transmon system. In other words, we adopt fixed-frequency transmons to suppress frequency fluctuations, known as dephasing noise, due to flux noise. Each qubit is detuned by approximately 100 MHz to avoid qubit-qubit modal interactions. This level of detuning reduces interference among modes and allows for two-qubit gates, such as the CNOT. The resonator is detuned to approximately 2 GHz above the qubit frequencies for a similar reason.

The Hamiltonian of the three-transmon system, $\hat{H}$, is defined according to its configuration. In Figure \ref{fig:transmon_systems}d, the island of each transmon qubit is capacitively connected to a single resonator coupler. Thus, the Hamiltonian consists of the Hamiltonian of each transmon qubit, $\hat{H}_{T,i}$, the resonator $\hat{H}_R$, and the transmon--resonator interaction $\hat{H}_I$.

\begin{equation}
  \hat{H}(t) = \hat{H}_{R} + \sum_{i=0}^{2}{\hat{H}_{T,i}(t)} + \hat{H}_{I}
\end{equation}
$\hat{H}_{T,i}$ represents the Hamiltonian of the $i$-th transmon.The total transmon Hamiltonian is expressed as $\hat{H}_T = \sum_{i=0}^{2}{\hat{H}_{T,i}}$.
The specifications of the transmons, such as their energy levels and anharmonicity, are characterized by the charging energy $E_{C,i}$ and the Josephson energy $E_{J,i}$\cite{Nakamura99,Koch07}.
\begin{equation}
  \hat{H}_{T,i}(t) = 4E_{C,i}(\hat{n}_i - n_{g,i}(t))^2 - E_{J,i} \cos\hat{\phi}_i
\end{equation}
Here, we utilize $\hbar=1$, so that time is expressed in units of ns and energy in units of GHz.

The charging energy depends on the gate capacitance and the Josephson junction (JJ) of the transmon. This value not only describes the transmon's susceptibility to charge but also its anharmonicity\cite{Koch07}. Hence, the charging energy indicates the transmon's sensitivity to an external microwave source. The Josephson energy is proportional to the critical current of the JJ. If the Josephson energy increases, the standard deviation of the eigenfunction of the state of the extra Cooper pair, represented by the number of Cooper pairs, also increases. This means that the transmon is less susceptible to charge fluctuations. Since the Josephson energy of a transmon is generally much larger than its charging energy, the Josephson energy primarily determines the transmon's energy level. The charge number operator, $\hat{n}_i$, corresponds to the number of extra Cooper pairs on the island for the ith transmon. The tunneling operator, $\cos\hat{\phi}_i$, describes the tunneling effect of a Cooper pair between the island and the reservoir for the i-th transmon. The gate offset number, $n_{g,i}(t)$, indicates the dimensionless microwave voltage pulse that drives the $i$-th transmon qubit. The resonator is modeled as an LC circuit with a resonance frequency $\omega_R$, which can be described by a harmonic oscillator.
\begin{equation}
  \hat{H}_{R} = \omega_R \hat{a}^\dagger \hat{a}
\end{equation}
$\hat{a}(\hat{a}^\dagger)$ is the annihilation(creation) operator of the resonator. The interaction Hamiltonian is expressed by the energy exchange between the transmon and the resonator.
\begin{equation}
  \hat{H}_I = \sum_{i=0}^{2}{G_{i}(\hat{a} + \hat{a}^\dagger) \hat{n}_i}
\end{equation}
The coupling energy $G_i$ between the $i$-th transmon and the resonator is determined by the capacitance of the coupling capacitor.

We choose an appropriate basis for simulating the three-transmon system. We employ an eigenenergy basis for each device.
\begin{align}
  \hat{H}_{T,i}(n_{g,i}=0)\ket{m_i}_{T,i} & = E_{m_i,i}\ket{m_i}_{T,i} \\
  \hat{H}_{R}\ket{k}_R                    & = E_{k}\ket{k}_R
\end{align}
$\ket{m_i}_{T,i}$ is the eigenstate corresponding to the $m$-th eigenenergy  $E_{m_i,i}$ for $i$-th transmon in idle state($n_{g,i}=0$). The eigenstate $\ket{k}_{R}$ corresponding to the eigenenergy $E_k$ is Fock state of the resonator.
. A qubit state can exist in a computational basis or at higher levels in real hardware. The latter case is known as a leakage error. Although transitions to higher levels are typically undesirable, they play a crucial role in implementing a CNOT gate based on a CR pulse\cite{Magesan20}.
Therefore, in our simulation, we included both the computational basis states $\{\ket{0}_{T,i},\ket{1}_{T,i}\}$ and the leakage $\{\ket{2}_{T,i},\ket{3}_{T,i}\}$ of transmon qubits in the simulation to implement accurate gates. The resonator coupler should not absorb any energy while mediating the qubits. The operation is considered successful if $\ket{0}_R$ is measured. However, if the measurement result indicates one of the states $\{\ket{1}_R, \ket{2}_R, \ket{3}_R\}$, an error has occurred in which the coupler has absorbed energy from the qubit. Therefore, the simulation basis $\mathcal{S}$ of the three-transmon system and the computational basis $\mathcal{C}$ which is the subset of $\mathcal{S}$ is defined as,
\begin{align}
  \mathcal{S} & = \left\{
  \ket{\Psi} = \ket{k}_R\ket{m_0}_{T,0}\ket{m_1}_{T,1}\ket{m_2}_{T,2}|
  k,m_i = 0,1,2,3\text{ for }i=0,1,2
  \right\},               \\
  \mathcal{C} & = \left\{
  \ket{\Psi} = \ket{k}_R\ket{m_0}_{T,0}\ket{m_1}_{T,1}\ket{m_2}_{T,2}|
  k=0, m_i=0,1\text{ for }i=0,1,2
  \right\}.
\end{align}

\subsection*{Pulse Protocol for CNOT Gates}\label{subsec:pulse}
\begin{figure}[t!]
  \centering
  \includegraphics[width=\linewidth]{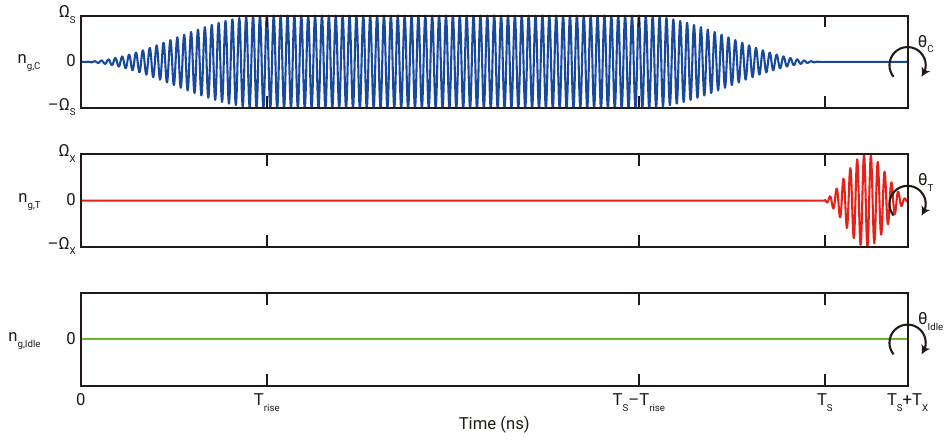}
  \caption{
    Pulse protocol for CNOT gate implemented in the three-transmon system. The Blue(red) line denotes the offset number $n_g(t)$ of the control(target) qubit. The idle qubit is at rest during the CNOT gate. $n_g(t)$ of the idle qubit is illustrated as the green line.
    The CR(auxiliary) pulse is applied by the gate time $T_{S(X)}$. $T_\text{rise}$ refers to the rising time of the CR pulse, which is the time it takes from zero pulse to the plateau of the envelope. The rotating arrows at the time $T_S+T_X$ represent VZ gates. The rotating angle of each qubit is $\theta_\text{C}$, $\theta_\text{T}$, and $\theta_\text{Idle}$.
  }
  \label{fig:pulseProtocol}
\end{figure}
We apply microwave voltage pulses to transmon qubits to perform quantum gates in the three-transmon system, focusing specifically on implementing CNOT gates rather than trivial single-qubit gates. An arbitrary single-qubit gate can be decomposed into a combination of physical $R_X$ gates and non-physical VZ gates\cite{McKay17}. The $R_X$ gate is easily implemented by applying a microwave pulse with a frequency that matches the target qubit frequency with high fidelity. CNOT gates in the three-transmon system are implemented using CR pulses\cite{Chow11,Kirchhoff18,Rigetti10}. The CR pulse induces a conditional resonance effect, causing the state of the target qubit to rotate based on the state of the control qubit.

Figure \ref{fig:pulseProtocol} illustrates the pulse protocol for implementing the CNOT gate in the three-transmon system, with detailed pulse information provided in the Methods section. The protocol comprises three steps: 1. Applying a CR pulse to the control qubit, 2. Applying an auxiliary pulse to the target qubit, and 3. Applying VZ gates to each qubit. The first step is central to the CNOT pulse protocol. During this step, the CR pulse frequency is tuned near the target qubit frequency, and pulse parameters are adjusted to construct the CNOT gate skeleton, involving two-qubit rotations such as $R_{ZX}$. It is crucial to detune the qubit frequency to prevent X rotation of the control qubit during this phase. The second step, following the CR pulse, involves rotating the target qubit state. The initial phase of the auxiliary pulse is a critical parameter that determines the rotational axis of the state on the Bloch sphere. In this study, we used the derivative removal by adiabatic gate (DRAG)\cite{Motzoi09} technique to suppress leakage errors. The DRAG envelope coefficient was set to $0.5/E_{C,T}$, where $E_{C,T}$ is the charging energy of the target qubit. The final step involves frame rotation without physical pulses, compensating for the relative phase of the CNOT gate. Note that this step only addresses single-qubit rotations. If the gate involves improper two-qubit rotations, such as $R_{ZZ}$, the fidelity of the CNOT gate is not improved by this final step. The total gate time for the CNOT gate is the sum of the CR pulse time $T_S$ and the auxiliary pulse time $T_X$. Figure \ref{fig:pulseProtocol} shows that the majority of the gate time is consumed by the CR pulse. Thus, we aimed to minimize the CR pulse time to reduce the overall gate time. Excessively increasing the amplitude of the CR pulse to shorten gate time can cause the transmon dynamics to deviate from typical cross-resonance effects, leading to an incorrect mapping of the initial two-qubit state. Meanwhile, the rising time $T_\text{rise}$ involves an adiabatic process in the control qubit, and setting an appropriate rising time is crucial to avoid unnecessary nonadiabaticity\cite{Nguyen22}.

\subsection*{Quantum Gate Optimization}\label{subsec:gateOpt}
\begin{table}[b!]
  \centering
  \caption{The specification of hardware in the three-transmon system. The unit of energy is GHz.}
  \label{table:hardware_spec}
  \begin{tabularx}{1.0\textwidth}{
      >{\centering\arraybackslash}X
      >{\centering\arraybackslash}X
      >{\centering\arraybackslash}X
      >{\centering\arraybackslash}X
      >{\centering\arraybackslash}X
    }
    \hline
    \multicolumn{1}{l}{} & $R$ & $T_0$   & $T_1$   & $T_2$   \\ \hline
    $\omega_R / 2\pi$    & 7.0 & -       & -       & -       \\
    $E_{C,i}/2\pi$       & -   & 0.30783 & 0.30902 & 0.31040 \\
    $E_{J,i}/2\pi$       & -   & 11.914  & 11.412  & 10.993  \\
    $G_{i}/2\pi$         & -   & 0.07    & 0.07    & 0.07    \\ \hline
  \end{tabularx}
\end{table}
We obtain the CNOT gate $\hat{\mathcal{U}} = \hat{\mathcal{Z}}\hat{\mathcal{U}}_\text{pulse}$ based on the Hamiltonian, as described in the previous sections.  The designed gate consists of the time-evolution operator $\hat{\mathcal{U}}_\text{pulse}$ and the VZ gate $\hat{\mathcal{Z}}$. The operator $\hat{\mathcal{U}}_\text{pulse}$ describes the evolution of the three-transmon system driven by the voltage pulses $n_{g,i}(t)$ applied over the total pulse time $T$,
\begin{equation}
  \hat{\mathcal{U}}_\text{pulse}
  = \hat{\mathcal{T}} \exp\left(-i\int_{0}^{T}{dt \hat{H}(t)}\right)
  = \hat{\mathcal{T}}\prod_{n=1}^{N}{e^{\left(-i\tau \hat{H}((n+1/2)\tau)\right)}},
\end{equation}
where $\hat{\mathcal{T}}$ denotes the time-ordering operator.
The integral term in the time-evolution operator is generally expressed as a Dyson series.
We divide the total evolution time $T$ into $N$ time intervals for the calculation, treating the Hamiltonian as a constant operator within each interval. In this work, we set the time interval $\tau=T/N$ to 1 ps to accurately capture the system's dynamics with minimal distortion. The exponential terms can be calculated using exact diagonalization. However, due to the high computational cost of exact diagonalization, we used the Suzuki-Trotter algorithm\cite{Suzuki85} to perform these calculations more efficiently.
\begin{equation}
  e^{\left(-i\tau \hat{H}(\tilde{t}_n)\right)} \simeq
  e^{\left(-i\tau \hat{H}_0/2\right)}
  Ve^{\left(-i\tau \hat{\Lambda}(\tilde{t}_n)\right)}V^\dagger
  e^{\left(-i\tau \hat{H}_0/2\right)}
\end{equation}
The Hamiltonian $\hat{H}(t) = \hat{H}_0 + \hat{H}_1(t)$ is divided into two parts: the diagonal part $\hat{H}_0$ and the non-diagonal part $\hat{H}_1(t)$ which can be easily diagonalized to $\hat{\Lambda}(t) = \hat{V}^\dagger H_1(t)\hat{V}$. Here $\tilde{t}_n$ is equal to $(n+1/2)\tau$. The local error bound of the approximation is proportional to $\tau^3$\cite{Raedt87}. The $\hat{\mathcal{Z}}$ operator is expressed as parallel $\hat{R}_{Z,i}(\theta_i)$ gates, which are not physical gates but represent system frame rotations, known as the VZ gates. These $Z$ rotations correct the relative phase of each qubit after applying the pulses.
\begin{equation}
  \hat{\mathcal{Z}} = \bigotimes_{i=0}^{2} \hat{R}_{Z,i}(\theta_i)
\end{equation}

\begin{table*}[t!]
  \centering
  \caption{
    The pulse parameters of the CNOT gate and the average fidelity in the three-transmon system.
  }
  \label{table:three-transmon_gate}
  \begin{tabularx}{1.0\textwidth}{
      >{\centering\arraybackslash}X
      >{\centering\arraybackslash}X
      >{\centering\arraybackslash}X
      >{\centering\arraybackslash}X
      >{\centering\arraybackslash}X
      >{\centering\arraybackslash}X
      >{\centering\arraybackslash}X
    }
    \hline
    Gate         & $f_1$ [GHz] & $f_2$ [GHz] & $T_X$ [ns] & $T_S$ [ns] & $\Omega_X$ & $S_q$, $\Omega_S$ \\ \hline
    $\CNOT_{01}$ & 4.9783      & 4.9783      & 10         & 130        & 0.0055     & $S_2$, 0.07       \\
    $\CNOT_{10}$ & 5.0851      & 5.0851      & 10         & 180        & 0.025      & $S_2$, 0.045      \\
    $\CNOT_{12}$ & 4.8895      & 4.8895      & 10         & 120        & 0.01       & $S_1$, 0.05       \\
    $\CNOT_{21}$ & 4.9783      & 4.9783      & 10         & 170        & 0.025      & $S_1$, 0.05       \\
    $\CNOT_{20}$ & 5.0851      & 5.0841      & 8.5        & 190        & 0.035      & $S_1$, 0.08       \\
    $\CNOT_{02}$ & 4.8895      & 4.8895      & 10         & 110        & 0.016      & $S_1$, 0.04       \\ \hline \hline
    $\varrho$    & $\gamma_1$  & $\gamma_2$  & $\theta_0$ & $\theta_1$ & $\theta_2$ & $F$               \\ \hline
    0.25         & 0           & 2.2007      & 0.6959     & 0          & 0.1001     & 0.9946            \\
    0.25         & 0           & -2.0420     & 0          & 2.5133     & 0          & 0.9913            \\
    0.3          & 0           & -1.5708     & 0          & 1.4451     & 0          & 0.9640            \\
    0.3          & 0           & -1.8850     & 0.1        & -0.1       & -1.5708    & 0.9884            \\
    0.2          & 0           & -1.2566     & 0.1571     & 0          & 2.6704     & 0.9938            \\
    0.3          & 0           & -2.3876     & -0.6283    & 0.1257     & 0.1571     & 0.9713            \\ \hline
  \end{tabularx}
\end{table*}
We estimate the performance of the quantum gate using the average fidelity $F$ for $M=10^4$ arbitrary pure states $\ket{\psi_j}$\cite{Nielsen02}.
\begin{equation}
  F = \frac{1}{M}\sum_{j=1}^{M} F_{\psi,j}
\end{equation}
The fidelity $F_{\psi,j}$ for a pure state $\ket{\psi_j}$ is defined as $F_{\psi,j} = |\bra{\psi_j}U^\dagger \mathcal{U}\ket{\psi_j}|$. $U$ denotes the target gate to be implemented. Note that we only considered transitions between computational basis states, meaning $U$ is independent of transitions related to leakage states.

Finally, we optimized the quantum gate as a function of the pulse parameters. To achieve this, we set up a simple optimization problem for the infidelity, $1-F$.
\begin{equation}
  \text{Minimize }1-F(\mathcal{U})\text{ for given }U.
\end{equation}
We utilized the Nelder-Mead algorithm\cite{Nelder65} to solve this problem because it is an optimization problem without constraints and involves $\mathcal{U}$, for which the gradient is not obtainable in practice. Note that $F(\mathcal{U})$ depends on the pulse parameters and hardware specifications. Here, the hardware specifications are not explicitly included in the optimization problem. Proper hardware configuration is crucial for implementing high-fidelity quantum gates. In other words, using incorrect specifications results in a higher lower bound for the infidelity.

\subsection*{Design of the Three-transmon System}\label{subsec:three-transmonSystem}

\begin{figure*}[t!]
  \centering
  \includegraphics*[width=0.9\textwidth]{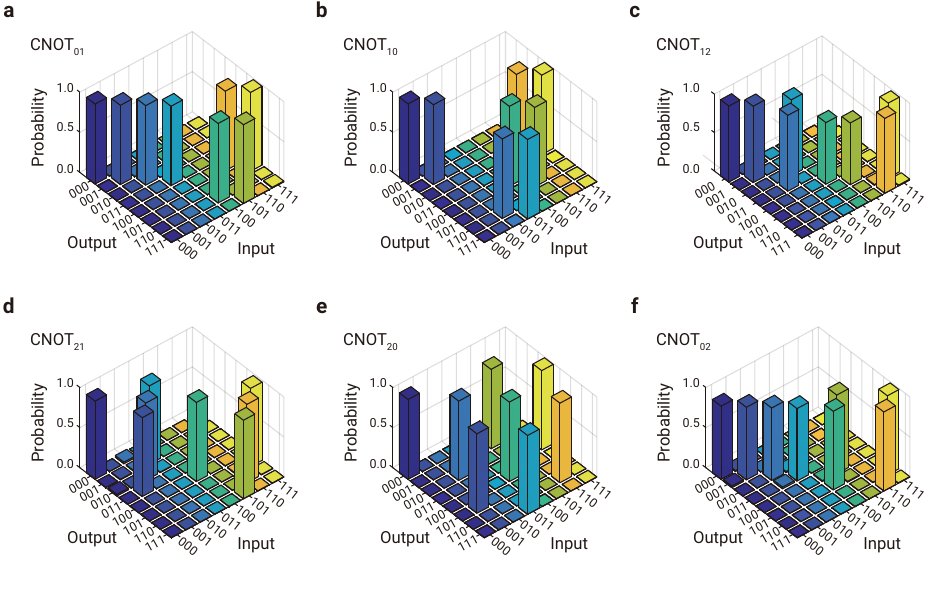}
  \caption{The success probability of CNOT gates for the three-qubit system in the computational basis.
    The average of the gate success probability of
    (a)$\CNOT_{01}$,
    (b)$\CNOT_{10}$,
    (c)$\CNOT_{12}$,
    (d)$\CNOT_{21}$,
    (e)$\CNOT_{20}$, and
    (f)$\CNOT_{02}$
    are 0.9975, 0.9967, 0.9681, 0.9882, 0.9928, 0.9720 for the basis states respectively.
  }
  \label{fig:gate_success_probabilities}
\end{figure*}
We present the design of a three-transmon system and propose the hardware specifications and pulse parameters necessary for performing CNOT gates within this system. We demonstrate the performance and operation of the CNOT gate in our new architecture by presenting the average fidelity, gate success probability for the basis states, and the trajectory of the Bloch vector. Table \ref{table:hardware_spec} lists the hardware specifications for the three-transmon system simulated in this study. We base these specifications on the IBM hardware, \texttt{ibm\_sherbrooke}, which can be replicated in real hardware. The qubit frequencies of each qubit $\omega_i$ is set by $\omega_0 = 5.0851$ GHz, $\omega_1 = 4.9783$ GHz, and $\omega_2 = 4.8895$ GHz. We confirm that each qubit's frequency was detuned by approximately 100-200 MHz, which is within the typical microwave frequency range. With these frequencies, we anticipate that a CNOT gate can be effectively implemented based on the cross-resonance effect in the three-transmon system.

Table \ref{table:three-transmon_gate} lists the pulse parameters and the average fidelity of the CNOT gates in the three-transmon system. Our system can implement CNOT gates for the following qubit pairs: $\CNOT_{01}$, $\CNOT_{10}$, $\CNOT_{12}$, $\CNOT_{21}$, $\CNOT_{20}$, and $\CNOT_{02}$. The frequencies of the pulses are normally tuned to the target qubit's frequency, except in the case of $\CNOT_{20}$. For this gate, the detuned auxiliary pulse complements $\mathcal{U}_\text{pulse}$ by providing additional single-qubit rotations, such as $R_Y$. These extra rotations help mitigate unnecessary qubit-qubit interactions, thereby enhancing gate fidelity. The gate time for the CNOT gates is less than 200 ns, indicating that our CNOT gates offer comparable fidelity and shorter gate times compared to CNOT gates based on echoed cross-resonance (ECR) gate\cite{Sheldon16}. We suggest that the reduced gate time of our CNOT gates can improve practical quantum circuit depth. The parameters $q$ and $\varrho$ refer to the shape of the envelope for the CR pulse and the ratio of the rising time to pulse time, respectively. These parameters are related to the adiabatic processes.

Figure \ref{fig:gate_success_probabilities} shows the gate success probabilities for the CNOT gates acted on the computational basis states. The success probability, which measures the likelihood that the basis state is correctly mapped by the ideal CNOT gate, reflects the quality of $\mathcal{U}_\text{pulse}$ and includes the effects of both the CR and auxiliary pulses. On average, the success probabilities for all CNOT gates in the three-transmon system exceed 0.98. Notably, at least one CNOT gate between any two arbitrary qubits in the system can be implemented with a success probability greater than 0.988. These results demonstrate that high-fidelity CNOT gates can be effectively implemented in a three-transmon system.

\begin{figure}[t]
  \centering
  \includegraphics[width=\linewidth]{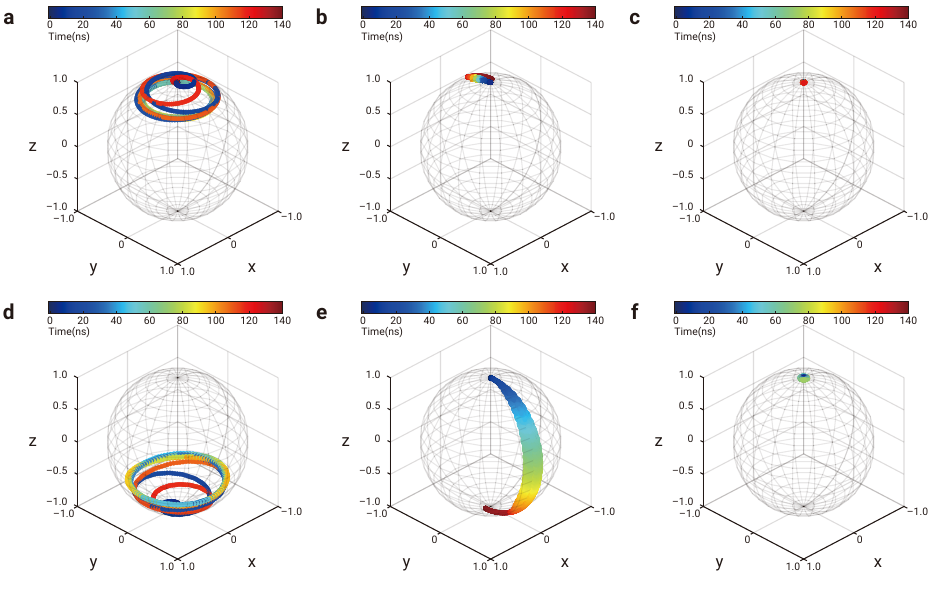}
  \caption{
    Trajectory of the Bloch vector for each qubit during $\CNOT_{01}$.
    The passage of time flows from blue to red as indicated by the colorbar.
    The Bloch vector of `Transmon 0'(a, d), `Transmon 1'(b, e), and `Transmon 2'(c, f) are illustrated in the case of the initial state $\ket{000}$(a, b, c) and $\ket{100}$(d, e, f).
  }
  \label{fig:CNOT01_Bloch}
\end{figure}
We investigate the expected values of all the Pauli operators for each qubit while applying the designed CNOT gates. This provide a comprehensive understanding of the CNOT operation, based on the trajectories of the Bloch vectors, which correspond to the time evolution of these expectation values. Figure \ref{fig:CNOT01_Bloch} illustrates the Bloch vector trajectories for each qubit during the application of $\CNOT_{01}$. Notably, the quantum state of the transmon not involved in the gate remained unchanged (Figures \ref{fig:CNOT01_Bloch}c and \ref{fig:CNOT01_Bloch}f). Figure \ref{fig:CNOT01_Bloch} also shows that cross-resonance represents the most time-consuming part of the CNOT gate. The qubit states during the gate are monitored throughout the CR pulse. During the rising and falling intervals of the pulse, the control qubit undergoes an adiabatic process (Figures \ref{fig:CNOT01_Bloch}a and \ref{fig:CNOT01_Bloch}d). In this process, the Bloch vector of the control qubit moves from the pole of the Bloch sphere toward the equator and then returns to its original position, indicating that the pulse envelope was properly designed. At the plateau stage, the control qubit only experiences a physical $Z$ rotation resulting from the detuned pulse frequency from the qubit frequency of the control qubit. The target qubit undergoes $X$ (Figure \ref{fig:CNOT01_Bloch}b) or $-X$ rotation (Figure \ref{fig:CNOT01_Bloch}e) according to the control qubit state. We do not approximate the time evolutions using the rotating-wave approximation. Therefore, rapid oscillations can be observed from the evolutions. The target qubit state reaches the opposite pole when an auxiliary pulse is applied after the CR pulse. We rechecked the operation of the designed CNOT gates operated, as shown in Figure \ref{fig:CNOT01_Bloch}.

\section*{Discussion}\label{sec:discussion}

This paper proposes a new structure based on a three-transmon system. Our design features three fixed-frequency transmon qubits and a single-resonator coupler that mediates interactions among all the qubits. Each transmon qubit has its own independent microwave drive lines for single- and two-qubit operations. With the use of CR pulses, the average fidelity of all CNOT gates in the three-transmon system exceeds 0.98. Given the low cost of single-qubit gates, CNOT gates with a fidelity of at least 0.988 can be achieved. These results suggest that high-fidelity two-qubit gates can be implemented in a system where a single resonator coupler mediates interactions among three transmon qubits. Inspired by the potential of using a tunable transmon system in regions less susceptible to flux noise to control qubit frequency detuning caused by long-term drift, we also explore the possibility of constructing a three-transmon system using tunable transmons. Additionally, we investigate configurations where a single-resonator coupler mediates multiple transmon qubits and consider the potential for structures incorporating new elements beyond traditional transmon-based quantum computers.

\section*{Methods}\label{sec:methods}

\begin{table}[t]
  \centering
  \caption{The specification of hardware in the two-transmon system. The unit of energy is GHz.}
  \label{table:2Q_hardware_spec}
  \begin{tabularx}{1.0\textwidth}{
      >{\centering\arraybackslash}X
      >{\centering\arraybackslash}X
      >{\centering\arraybackslash}X
      >{\centering\arraybackslash}X
    }
    \hline
    \multicolumn{1}{l}{} & $R$ & $T_0$  & $T_1$   \\ \hline
    $\omega/2\pi$        & 7.0 & -      & -       \\
    $E_{C,i}/2\pi$       & -   & 0.3461 & 0.3421  \\
    $E_{J,i}/2\pi$       & -   & 9.9178 & 10.9781 \\
    $G_{i}/2\pi$         & -   & 0.07   & 0.07    \\ \hline
  \end{tabularx}
\end{table}

% \subsection*{Design of Pulses}\label{subsec:pulseProtocol}
In this section, we describe the pulse protocol for implementing CNOT gates. The microwave pulse is represented by the gate offset number $n_{g,i}(t)$ in the Hamiltonian of the $i$-th transmon. While multiple pulses can generally be applied to a qubit, we apply at most one pulse per qubit to implement all gates.
\begin{equation}
  n_{g,i}(t) = \Omega_i(t)\cos(2\pi f_i t - \gamma_i)
\end{equation}
$\Omega_i(t)$ denotes the envelope of the pulse, which includes both in-phase and quadrature components. The frequency of the pulse is denoted as $f_i$. The initial phase of the pulse $\gamma_i$ determines the axis of rotation on the Bloch sphere.

We use two types of envelope: $\Omega_G(t)$ and $\Omega_{S,q}(t)$.
\begin{equation}
  \Omega_G (t) = \begin{cases}
    \Omega_X \frac{\exp\left\{\left.-\left(t-T_X/2\right)^2 / 2\sigma^2\right.\right\}
    - \exp\left\{\left.-T_X^2/8\sigma^2\right.\right\}}{1-\exp\left\{\left.-T_X^2/8\sigma^2\right.\right\}} & (0 \leq t \leq T_X) \\
    0                                                                                                       & \text{Otherwise}
  \end{cases}
\end{equation}
\begin{equation}
  \Omega_{S,q}(t) = \begin{cases}
    \Omega_S S_q(t)     & (0 \leq t < T_\text{rise})                             \\
    \Omega_S            & (T_\text{rise} \leq t < T_\text{rise} + T_S)           \\
    \Omega_S S_q(t-T_S) & (T_\text{rise} + T_S \leq t \leq 2T_\text{rise} + T_S) \\
    0                   & \text{Otherwise}
  \end{cases}
\end{equation}
$\Omega_G(t)$ is a Gaussian envelope used for single-qubit rotations such as $R_X$. It is characterized by the pulse time $T_X$, the amplitude $\Omega_X$, and the pulse width $\sigma = T_X/4$. The other envelope $\Omega_{S,q}(t)$ is a sinusoidal flat-top envelope used for CNOT gate. The subscript $q$ indicates the shape function $S_q(t)=\sin^q(\pi t/ 2T_{\text{rise}})$. This envelope is characterized by the pulse time $T_S$, the amplitude $\Omega_S$, and the rising ratio $\varrho = T_\text{rise}/T_S$. The rising time $T_\text{rise}$ represents the time required to reach the plateau.

% \subsection*{Asymmetric CNOT}

\begin{table*}[t!]
  \centering
  \caption{
    The pulse parameters of symmetric CNOT gate and the average fidelity in the two-transmon system.
  }
  \label{table:two-transmon_gate_sym}
  \begin{tabularx}{1.0\textwidth}{
      >{\centering\arraybackslash}X
      >{\centering\arraybackslash}X
      >{\centering\arraybackslash}X
      >{\centering\arraybackslash}X
      >{\centering\arraybackslash}X
      >{\centering\arraybackslash}X
    }
    \hline
    Gate                         & $f_C$ [GHz]    & $f_T$ [GHz]           & $T_{X,C}$ [ns] & $T_{X,T}$ [ns]             & $T_{\rm CR \it}$ [ns] \\ \hline
    $\CNOT_{01}^{(\rm sym.\it)}$ & 4.8641         & 5.1108                & 13.095         & 13.095                     & 150                   \\
    $\CNOT_{01}^{(\rm sym.\it)}$ & 5.1108         & 4.8641                & 13.095         & 13.095                     & 60                    \\ \hline\hline
    $\Omega_{X,C}$               & $\Omega_{X,T}$ & $\Omega_{\rm CR \it}$ & $\gamma_{1,C}$ & $\gamma_{2,C}$             & $\gamma_{3,C}$        \\ \hline
    0.026486                     & 0.01394        & 0.06                  & -1.5708        & 3.7699                     & 3.1416                \\
    0.027900                     & 0.01324        & 0.03                  & -1.5708        & 1.8850                     & 3.1416                \\ \hline \hline
    $\gamma_{4,C}$               & $\gamma_{1,T}$ & $\theta_0$            & $\theta_1$     & \multicolumn{2}{c}{$F$}                            \\ \hline
    0.1571                       & 0              & $0.8796$              & $-0.0314$      & \multicolumn{2}{c}{0.9921}                         \\
    -0.9425                      & 0              & $0$                   & $-0.2199$      & \multicolumn{2}{c}{0.9910}                         \\ \hline
  \end{tabularx}
\end{table*}
\begin{table*}[t!]
  \centering
  \caption{
    The pulse parameters of asymmetric CNOT gate and the average fidelity in the two-transmon system.
  }
  \label{table:two-transmon_gate_asym}
  \begin{tabularx}{1.0\textwidth}{
      >{\centering\arraybackslash}X
      >{\centering\arraybackslash}X
      >{\centering\arraybackslash}X
      >{\centering\arraybackslash}X
      >{\centering\arraybackslash}X
      >{\centering\arraybackslash}X
      >{\centering\arraybackslash}X
    }
    \hline
    Gate                          & $f_1$ [GHz] & $f_2$ [GHz] & $T_X$ [ns] & $T_S$ [ns] & $\Omega_X$                 & $S_q$, $\Omega_S$ \\ \hline
    $\CNOT_{01}^{(\rm asym.\it)}$ & 5.1108      & 5.1108      & 10         & 200        & 0.029                      & $S_1$, 0.1        \\
    $\CNOT_{10}^{(\rm asym.\it)}$ & 4.8641      & 4.8641      & 10         & 129        & 0.012                      & $S_2$, 0.05       \\ \hline \hline
    $\varrho$                     & $\gamma_1$  & $\gamma_2$  & $\theta_0$ & $\theta_1$ & \multicolumn{2}{c}{$F$}                        \\ \hline
    0.16                          & 0           & -0.9425     & $1.5394$   & $-0.0628$  & \multicolumn{2}{c}{0.9979}                     \\
    0.25                          & 0           & 0.3142      & $-0.0628$  & $2.6389$   & \multicolumn{2}{c}{0.9977}                     \\ \hline
  \end{tabularx}
\end{table*}
We introduce an asymmetric CNOT and compare its performance with that of a symmetric CNOT, also known as a CR2-type CNOT\cite{Willsch17} or ECR gate. For the asymmetric CNOT, we applied a CR pulse to the control qubit and an auxiliary pulse to the target qubit.
\begin{align}
  n_{g,C}(t) & = \Omega_S(t) \cos(2\pi f_1 t - \gamma_1)           \\
  n_{g,T}(t) & = \Omega_G(t-T_S) \cos(2\pi f_2 (t-T_S) - \gamma_2)
\end{align}
$C$($T$) denotes the index number of the control (target) qubits. Unlike the symmetric CNOT, which includes an echoed CR pulse that increases gate time, the asymmetric CNOT does not use an echo pulse. Consequently, we have shortened the gate time. This suggests that the asymmetric CNOT could perform better than the symmetric CNOT in real hardware, where relaxation and dephasing noises are present, due to its shorter interaction time during gate application. Additionally, the asymmetric CNOT has fewer optimization parameters compared to the symmetric CNOT. This smaller number of pulse parameters implies that the asymmetric CNOT requires less computational cost for optimization.

We examined both types of CNOT gates in a \textit{two-transmon system}, which consists of two transmon qubits and a resonator coupler that mediates capacitive interactions between the qubits. Table \ref{table:2Q_hardware_spec} lists the specifications of this two-transmon system. For the transmon qubits, we refer to the hardware specifications of \texttt{ibmq\_kolkata}. Tables \ref{table:two-transmon_gate_sym} and \ref{table:two-transmon_gate_asym} provide the pulse parameters and average fidelities of the symmetric and asymmetric CNOT gates, respectively, in this system. We highlight the following two points: 1. The gate times for the asymmetric CNOT gates($\sim$210 ns, 139 ns) are shorter than for the symmetric CNOT gates($\sim$326 ns, 206 ns). 2. The average fidelity of the asymmetric CNOT gates is higher than that of the symmetric CNOT gates. These results demonstrate that the asymmetric CNOT gate, incorporating the resonator coupler, is advantageous for implementing high-fidelity CNOT gates in transmon systems.

% \subsection*{Development of CNOT}

To properly implement and optimize the CNOT gates for the three-transmon system, we investigated the seed parameters of the CNOT gates in three steps. For the asymmetric CNOT gate, the parameter vector is a 12-dimensional real vector,
\begin{equation}
  (f_1, f_2, T_X, T_S, \Omega_X, \Omega_S,
  \varrho, \gamma_1, \gamma_2, \theta_0, \theta_1, \theta_2)^T.
\end{equation}
We first search the proper envelope $\Omega_{S,q}(t)$ of the CR pulse, which depends on the parameters, $f_1$, $T_S$, $q$, $\Omega_S$, $\varrho$, and $\gamma_1$. Note that the shape index $q$ is not variable for the pulse optimization but fixes the shape of the envelope $\Omega_{S,q}(t)$. The `sweet spot' is identified where the success probabilities of the CNOT gate, which is not fully developed, are nearly equal for the computational basis states. The success probabilities for the computational basis states suggest that the conditional resonance due to the CR pulse results in orthogonality of the resultant states. This process can be executed using a real quantum computer. The next step is to investigate the parameters of the auxiliary pulses: $f_2$, $T_X$, $\Omega_X$, and $\gamma_2$. In this step, we determine the conditions that locally maximize the success probabilities for the basis states. From the perspective of the Bloch vector, this step positions the target qubit at the poles of the Bloch sphere. Finally, we set up the rotation angles $\theta_i$ of the $\hat{R}_{Z,i}$ gates, which can be easily implemented, and estimate the average fidelity of the CNOT gate. If the fidelity is not sufficiently high, the parameter point is considered a local maximum and must be reset, requiring the entire process to be repeated. If an appropriate point is found, it is used as the initial point in the optimization problem.

% \bibliography{sample}

% \noindent LaTeX formats citations and references automatically using the bibliography records in your .bib file, which you can edit via the project menu. Use the cite command for an inline citation, e.g.  \cite{Hao:gidmaps:2014}.

% For data citations of datasets uploaded to e.g. \emph{figshare}, please use the \verb|howpublished| option in the bib entry to specify the platform and the link, as in the \verb|Hao:gidmaps:2014| example in the sample bibliography file.

\section*{Acknowledgements}

This work is supported by the Basic Science Research Program through the National Research Foundation of Korea (NRF) funded by the Ministry of Education, Science and Technology (NRF2022R1F1A1064459) and Creation of the Quantum Information Science R$\&$D Ecosystem (Grant No. 2022M3H3A106307411) through the National Research Foundation of Korea (NRF) funded by the Korean government (Ministry of Science and ICT).

\section*{Author contributions statement}

J. Kang and Y. Kwon conceived the main idea and conducted the experiments. J. Kang, C. Kim, Y. Kim, and Y.Kwon analyzed the results. J. Kang and Y. Kwon prepared the manuscript.  All authors reviewed the manuscript.

\section*{Competing interests}
The authors declare no competing interests.

\section*{Data availability}
The datasets generated and analysed during the current study are available in the `pulse\_parameters repository', \\
https://github.com/JeongsooKang/pulse\_parameters/tree/main.

% \section*{Additional information}

% To include, in this order: \textbf{Accession codes} (where applicable); \textbf{Competing interests} (mandatory statement). 

% The corresponding author is responsible for submitting a \href{http://www.nature.com/srep/policies/index.html#competing}{competing interests statement} on behalf of all authors of the paper. This statement must be included in the submitted article file.

% \begin{figure}[ht]
% \centering
% \includegraphics[width=\linewidth]{stream}
% \caption{Legend (350 words max). Example legend text.}
% \label{fig:stream}
% \end{figure}

% \begin{table}[ht]
% \centering
% \begin{tabular}{|l|l|l|}
% \hline
% Condition & n & p \\
% \hline
% A & 5 & 0.1 \\
% \hline
% B & 10 & 0.01 \\
% \hline
% \end{tabular}
% \caption{\label{tab:example}Legend (350 words max). Example legend text.}
% \end{table}

% Figures and tables can be referenced in LaTeX using the ref command, e.g. Figure \ref{fig:stream} and Table \ref{tab:example}.

\end{document}